\documentclass[10pt,letterpaper]{article}
\usepackage[top=0.85in,left=2.75in,footskip=0.75in,marginparwidth=2in]{geometry}

\usepackage[utf8]{inputenc}

\usepackage{cite}

\usepackage{csquotes}

\usepackage{mathrsfs}

\usepackage[official]{eurosym}

\usepackage{amsthm}

\theoremstyle{definition}
\newtheorem{definition}{Conjecture}[section]

\DeclareMathAlphabet{\mathpzc}{OT1}{pzc}{m}{it}

\usepackage{nameref,hyperref}

\usepackage[right]{lineno}

\usepackage{microtype}
\DisableLigatures[f]{encoding = *, family = * }

\raggedright
\usepackage{changepage}

\usepackage[aboveskip=1pt,labelfont=bf,labelsep=period,singlelinecheck=off]{caption}

\makeatletter
\renewcommand{\@biblabel}[1]{\quad#1.}
\makeatother

\usepackage{lastpage,fancyhdr,graphicx}
\usepackage{epstopdf}
\fancyheadoffset[L]{2.25in}
\fancyfootoffset[L]{2.25in}

\usepackage{color}

\definecolor{Gray}{gray}{.25}

\usepackage{graphicx}

\usepackage{sidecap}

\usepackage{wrapfig}
\usepackage[pscoord]{eso-pic}
\usepackage[fulladjust]{marginnote}
\reversemarginpar

\begin{document}
\vspace*{0.35in}

% title goes here:
\begin{flushleft}
{\Large
\textbf\newline{Disintermediation of Inter-Blockchain Transactions}
}
\newline
% authors go here:
\\
S. Matthew English,
Fabrizio Orlandi,
S\"{o}ren Auer
\\
\bigskip
Fraunhofer Institut f\"{u}r Intelligente Analyse und Informationssysteme (IAIS)\\
Rheinische Friedrich-Wilhelms-Universit\"{a}t Bonn
\\
\bigskip
\{english, orlandi, auer\}@cs.uni-bonn.de

\end{flushleft}

\section*{Abstract}
Different versions of peer-to-peer electronic cash exist as data represented by separate blockchains. 
Payments between such systems cannot be sent directly from one party to another without going through a financial institution.
Bitcoin provided part of the solution but its utility is limited to intra-blockchain transactions.
The benefits are lost if a trusted third party is required to execute \textit{inter}-blockchain transactions.
We propose a solution to the inter-blockchain transaction problem using the same fundamental principles of Bitcoin. 
The protocol is described by the \"{U}berledger framework, a hierarchical meta-blockchain layer that encapsulates information regarding the fidelity of peer-to-peer transaction facilitators. 

% now start line numbers
% \linenumbers

% the * after section prevents numbering
\section{Introduction}
In the world prior to the introduction of Bitcoin no peer-to-peer version of electronic cash had managed to solve the problem of double-spending.  
Nakamoto in \cite{satoshi2008bitcoin} described the state of the art whereby a common solution had been to \enquote{introduce a trusted central authority, or mint, that checks every transaction for double spending... The problem with this solution is that the fate of the entire money system depends on the company running the mint, with every transaction having to go through them, just like a bank}. 

Bitcoin is an effective mechanism to circumvent dependence on a trusted central authority however we have borne witness to the hydralike re-emergence of this phenomenon in the form of exchanges \textit{between} cryptocurrency systems.
For instance to transfer \euro{}100.00 of value between the cryptocurrency Bitcoin (BTC) and the cryptocurrency Ethereum (ETH) it is generally necessary to employ the services of an exchange such as Bitfinex\footnote{ \texttt{www.bitfinex.com}}, or another third party service.

In this work we propose a framework to utilize the same fundamental principles that Bitcoin used to disintermediate the exchange of value amongst participants on a common blockchain system to the problem of disintermediation between blockchains that are \textit{disjoint}. 

In the decentralized constellation of cryptocurrencies the focal points of centripetal force are digital currency exchangers. 
Centralized processes are susceptible to developing into single points of failure, a feature that resilient networks seek to minimize.
Digital currency exchangers provide primarily two services, they transfer value in digital currencies for fiat money, or different digital currencies.
We demonstrate that the interchange of value between different digital currencies is a task that can be performed with high fidelity in the absence of trusted third party exchangers through the use of a hierarchical meta-blockchain layer.

\begin{figure}
\centering
\includegraphics[width=\textwidth]{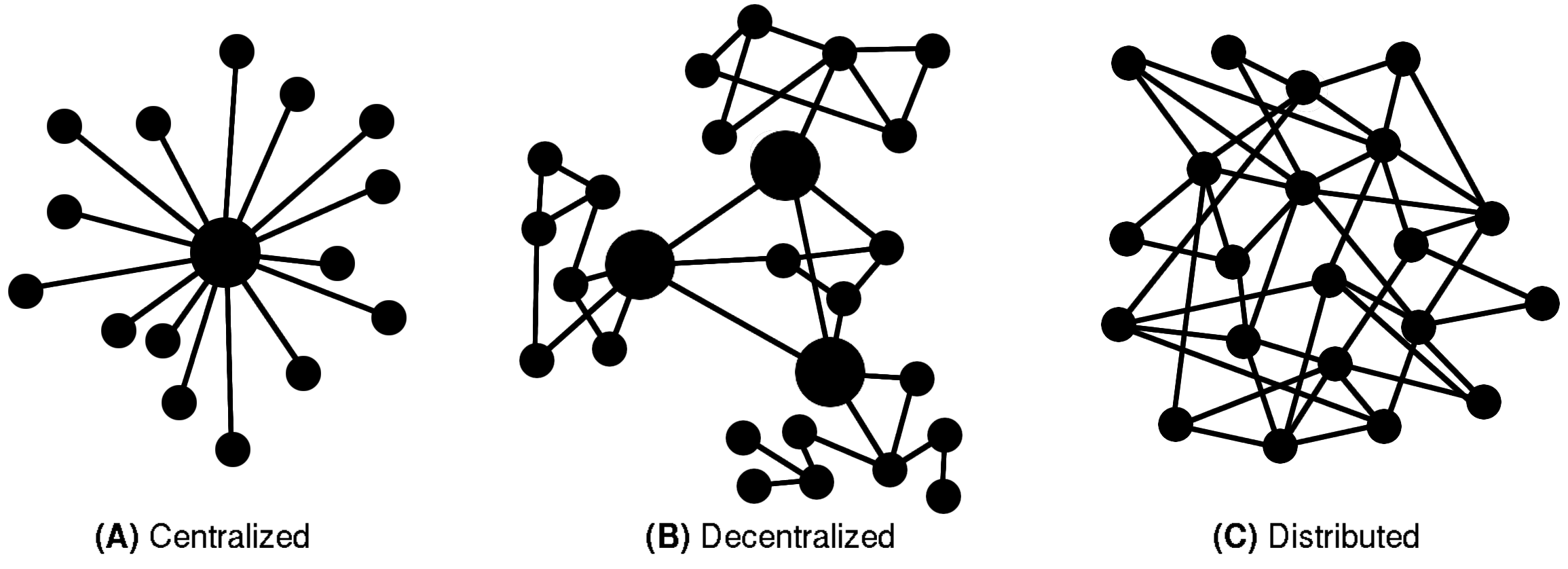}
\caption{Network Topologies}
\end{figure}

\section{The Exchange Model}

On the $2^{nd}$ of August 2016 a highly trafficked digital currency exchange reported a heist of \$72,000,000 USD \cite{baldwin_2016}. 
This is merely one episode in a history of criminal negligence or malfeasance going back to Mt. Gox (the infamous exchange responsible for the loss of \$460,000,000 in customer funds \cite{wire}) and beyond. 
Accordingly there is a strong interest in viable alternatives to transferring information, viz. value, between disparate blockchain systems independent from such central authorities.

Reliance on exchangers to move value between blockchain systems is a policy that suffers from the \enquote{inherent weaknesses of the trust based model} \cite{satoshi2008bitcoin}.
Cryptocurrency users who have interacted with an exchange will be familiar with the stringent regime of know your customer (KYC) and will have had direct experience with exchanges \enquote{wary of their customers, hassling them for more information than they would otherwise need}  \cite{satoshi2008bitcoin}. 
As originally envisioned by Nakamoto in \cite{satoshi2008bitcoin} our framework is engineered to \enquote{allow online payments to be sent directly from one party to another without going through a financial institution}. 

\begin{figure}
\centering
\includegraphics[width=0.7\textwidth]{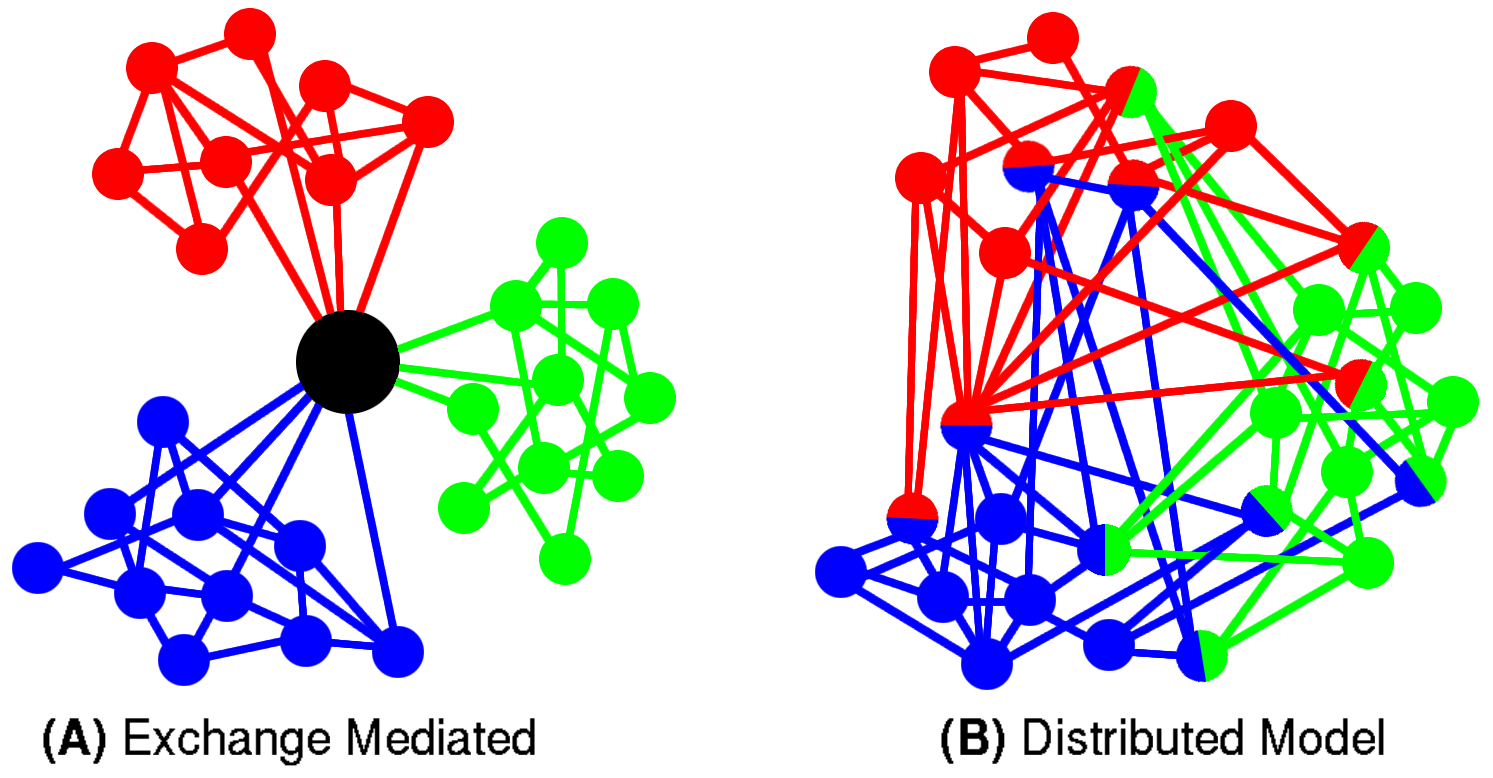}
\caption{Inter-Blockchain Value Transmission}
\end{figure}

\section{Related Work}

It has been shown by Thomas \& Schwartz in \cite{hope2016interledger} that protocols whereby a subset of network participants (with accounts on two distinct blockchains) is employed to act as \enquote{connectors and notaries}, i.e. facilitators of a transaction between ledgers, are feasible so long as the subset is sufficiently large to ensure that Byzantine actors can be readily identified ($\mathpzc{n}$ $\geq$ 3$\mathpzc{f}$+1). 

One design feature of the system described in \cite{hope2016interledger} is an ephemeral aggregation of transaction facilitators, such that \enquote{[facilitators] are organized in ad-hoc groups for each payment}.
This arrangement preserves the integrity of the funds involved in the transaction, either they are correctly allocated or the transaction is forfeit.
However information regarding the integrity of each node is not preserved in a publicly available repository of information, such as a blockchain, where it could be put to use in future transactions.  

In contrast to \cite{hope2016interledger} we seek to indelibly preserve all information regarding the successful (or unsuccessful) outcome of the transaction and the behaviour of constituent parties. 
The procedure sketched in Figure 3.

\section{Framework Architecture}

We introduce the \"{U}berledger framework\footnote{ \texttt{www.uberledger.io}}. It is a hierarchical blockchain model based on the following proposition:  

\theoremstyle{definition}
\begin{definition}{(\"{U}berledger)}
In the transference of value between two disjoint consensus networks the property of timestampedness cannot be preserved in the absence of an additional meta consensus network.
\end{definition}

Careful examination of Nakamoto's protocol in \cite{satoshi2008bitcoin} will yield timestampedness as the crux of the cryptocurrency system. 
We take timestampedness to mean that each transaction ($\mathpzc{T_i}$) is part of a linearly ordered list of actions ($\mathpzc{A_ix}$).
In our model consensus network is equivalent to blockchain, as defined by Kiayias et al. in \cite{garay2015bitcoin}. 

\begin{figure}
\centering
\includegraphics[width=0.7\textwidth]{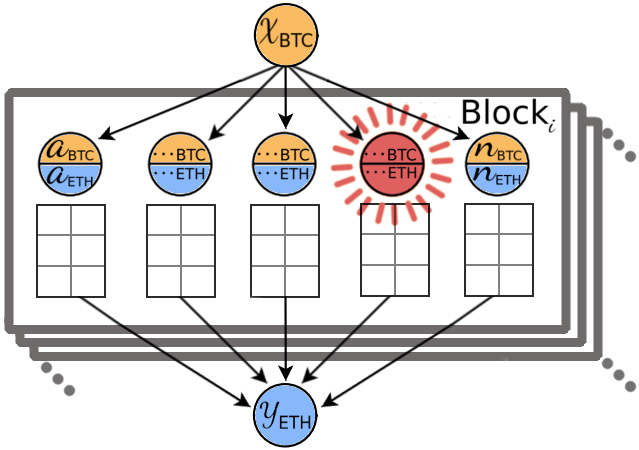}
\caption{Disintermediated Inter-Blockchain Transaction}
\end{figure}

\begin{figure}
\centering
\includegraphics[width=0.7\textwidth]{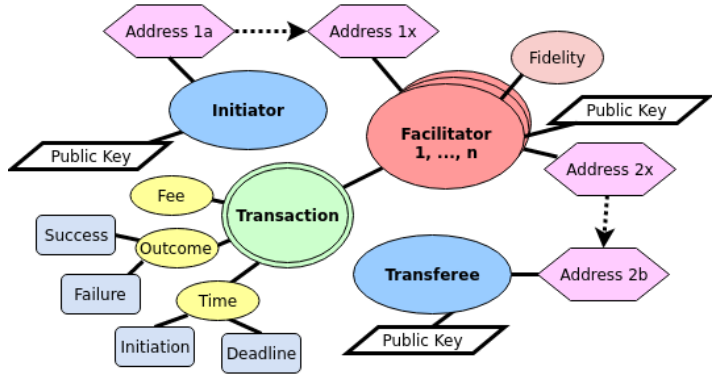}
\caption{Transaction Data Model}
\end{figure}

\section{Design Considerations}

\subsection{Incentive Structure}
The incentive structure that motivates the continued maintenance of a resilient blockchain is critical, as the undertaking is costly. 
Honest participants of the \"{U}berledger framework stand to be remunerated in proportion to their ability to attract transaction fees for their services.

\subsection{Data Representation}
Transactions are naturally represented in the form of a 3-tuple ($\mathpzc{P_1}$, $\mathpzc{a}$, $\mathpzc{P_2}$), where $\mathpzc{P}$ is a transacting party and $\mathpzc{a}$ is the article of trade. 
Accordingly we employ the RDF data model and encapsulate salient transaction information in the form of a graph, as demonstrated in Figure 4. 
To define data in the form of a graph it is necessary to employ a schema.
For this purpose we have adapted the blockchain ontology with dynamic extensibility (BLONDiE)\footnote{\texttt{www.github.com/EIS-Bonn/BLONDiE}}.

This model ensures that a wide range of disparate data resources, e.g. multiple accounts across different blockchains, unique username, reputation information, and cryptographic keys, are rendered in a standardized and universally accessible format adhering to the W3C principles of linked open data.

\subsection{Participant Evaluation}
Insofar as the integrity of nodes in the framework exists as a matter of public record we adapt the design considerations in \cite{kamvar2003eigentrust} to serve as fundamental benchmarks of our peer-to-peer reputation system, as such the protocol is: 
\begin{enumerate}
  \item Self-policing. 
  \item Anonymity maintaining.
  \item Negatively biased to newcomers.
  \item Minimal in overhead (i.e. computation, infrastructure, storage, and message complexity).
  \item Robust to malicious coordinated collectives.
\end{enumerate}

\section*{Discussion}
The balkanization of the cryptocurrency ecosystem is a phenomenon that is enforced by the business model of the exchanges that seek to exploit their control over channels into and out of different blockchains inhibiting those interested in experimentation on new platforms. Such a climate stifles the ability to assess innovative features and challenge one's understanding of novel techniques. 
Our framework is engineered to redress such toll roads on the highway of creative endeavour.
\"{U}berledger is an open source initiative that seeks to engender an environment of free creative development. 

\bibliography{library}

%This defines the bibliographies style. Search online for a list of available styles.
\bibliographystyle{abbrv}

\end{document}